\documentclass[a4paper,fleqn,oneside]{article}

\usepackage{mathtools,amsfonts,amssymb,amsthm,mathrsfs}
\usepackage{latexsym}
\usepackage{graphicx}
\usepackage{fixltx2e}
\usepackage{tensor}
\usepackage{caption}

\usepackage{indentfirst}
\usepackage[T1]{fontenc}

\usepackage{hyperref}

\usepackage[superscript]{cite}
\makeatletter
\def\@biblabel#1{$\null^{#1}$}
\makeatother
\newcommand{\beq}{\begin{equation}}
\newcommand{\eeq}{\end{equation}}

\newcommand{\bse}{\begin{subequations}}
\newcommand{\ese}{\end{subequations}}

\numberwithin{equation}{section}

\theoremstyle{plain}

\theoremstyle{definition}
\newtheorem{defn}{Definition}[section]

\theoremstyle{remark}

\newcommand{\ain}[1]{$\boldsymbol{#1}$}
\newcommand{\ainF}[1]{\boldsymbol{#1}}

\begin{document}
\title{\Large\textbf{Finding integrals and identities in the Newman Penrose formalism:\\ 
                    a comment on\\ 
\emph{``Petrov D vacuum spaces revisited: identities and invariant classification''} by B.S. Edgar 
                     et al.,\\ Class.\ Quantum Grav.\ \underline{26} (2009) 105022\\
      and on\\
      ``Type D vacuum solutions: a new intrinsic approarch'' by J.J. Ferrando \& J.A. S\'aez,\\
      Gen.\ Relativ.\ Gravit.\ (2014)\textbf{46}:1703 }}                    
\vspace{1cm}
\author{\textbf{Georgios O. Papadopoulos}\thanks{e-mail: gopapado@phys.uoa.gr}\\
\textit{National \& Kapodistrian University of Athens}\\
\textit{Department of Physics}\\
\textit{Nuclear \& Particle Physics Section}\\
\textit{Panepistimioupolis, Ilisia GR 157--71, Athens, Greece}\\}
\date{}
\maketitle

% % % % % % % % % % % % % % % % % % % % % % % % % % % % % % % % % % % % % % % % % % % % % % % % % % % % % % % %

\begin{abstract}
In 1969 W. Kinnersley\cite{Kinnersley}, using the \emph{Newman-Penrose}\cite{Penrose_Rindler} (NP) formalism, found all 
the Petrov type D, Ricci flat, solutions to the \emph{Einstein's Field Equations} (EFEs). Yet, in doing so ---as it seems--- 
he neglected two fundamental identities (or constraints) on four NP variables and Cartan invariants as 
well, namely $\tau\bar{\tau}-\pi\bar{\pi}=0$ and $\mu\bar{\rho}-\rho\bar{\mu}=0$. Since then, these identities 
have been constantly either overlooked or proven under special circumstances (e.g. electrovac 
solutions)\cite{Edgar_References}. It was only until 2009, when B.S. Edgar and his 
collaborators\cite{Edgar_el_al.}, by making an extended use of the 
Geroch-Held-Penrose (GHP) formalism\cite{Penrose_Rindler}, and of a computer algebra system (CAS), succeeded in 
proving those identities in the general case (i.e., within the Kinnerseley's assumptions). In that reference, 
it was ---rather indirectly--- implied that the results under consideration were provable only within the GHP 
formalism and thus the latter is the optimal tool towards the invariant classification and study of classes of 
solutions to the EFEs.\\
In 2014 there was a kind of response to that paper by J.J. Ferrando \& J.A. Sa\'ez\cite{Ferrando_Saez}. Using the 
tensorial algebra (of bivectors or 2-forms), and without the aid of a CAS, the authors proved the desired result
and they offer a much more refined and extended classification of the Petrov type D, Ricci flat, solutions. Never 
the less when someone reads that third work, although beautiful and conceptually simple, one has the feeling that the 
authors know in advance what they want to prove; something which is not always the case.

The goal of the present short work is to prove, through the specific example (i.e., the class of Ricci flat, 
Petrov type D geometries), that the original NP formalism, seen as an \emph{exterior differential system} (EDS)
suffices to provide the desired results ---thus commenting on the second cited paper--- not only without (in 
principle) the aid of a CAS, but also to obtain a new further (unknown until now) integral of the EDS ---thus
commenting on the third cited paper.

\vspace{0.3cm}
\noindent
\textbf{MSC-Class} (2010): 53B20, 53B30, 53A55, 53A45, 53B21, 53B50, 83C60, 83C15, 83C20 \\
\textbf{PACS-Codes} (2010): 02.40.-k, 04.20.-q, 04.20.Jb 
\end{abstract}

\section{The NP formalism as an Exterior Differential System (EDS)}
Let a pseudo-Riemannian space be described by the pair $(\mathcal{M},\mathbf{g})$, where $\mathcal{M}$ is a 4 dimensional, simply 
connected, Hausdorff, and $C^{\infty}$ manifold and $\mathbf{g}$ is a $C^{\infty}$ metric tensor field on it that is a non degenerate, 
covariant tensor field of order 2, with the property that at each point of $\mathcal{M}$ one can choose a frame of 4 real vectors 
$\{\ainF{e}_{1},\ldots,\ainF{e}_{4}\}$, such that $\mathbf{g}(\ainF{e}_{a},\ainF{e}_{b})=g_{ab}$ where \ain{g} (called 
\emph{frame metric}) is a constant symmetric matrix with prescribed signature.\\
The NP formalism\cite{Penrose_Rindler} can be thought of as a torsionless local geometry described by a (special) class of bundles of 
quasi orthonormal, semi complex frames. More precisely a linear, complex combination of the four real frame vectors is chosen such that it 
is
\beq
\ainF{e}_{a}=(\ainF{\Delta},\ainF{D},-\bar{\ainF{\delta}},-\ainF{\delta}), ~~~\bar{\ainF{\Delta}}=\ainF{\Delta}, 
~~\bar{\ainF{D}}=\ainF{D}
\eeq
where a bar over a symbol denotes complex conjugation, while
\beq
\mathbf{g}(\ainF{e}_{a},\ainF{e}_{b})=g_{ab}\equiv\eta_{ab}=
\begin{pmatrix}
0 & 1 & 0 & 0 \\
1 & 0 & 0 & 0  \\
0 & 0 & 0 & -1 \\
0 & 0 & -1 & 0
\end{pmatrix}
\eeq
and the coframe is defined via 
\beq
\ainF{e}^{}_{a} ~\lrcorner ~\ainF{\omega}^{b}_{}=\delta_{a}^{\phantom{1}b}
\eeq
as
\beq
\ainF{\omega}^{a}=(\ainF{l},\ainF{n},\ainF{m},\bar{\ainF{m}}), ~~~\bar{\ainF{l}}=\ainF{l}, ~~\bar{\ainF{n}}=\ainF{n}
\eeq
so that for any 0-form (i.e., a scalar) function $f$ it is\\ 
\beq
df=\ainF{e}^{}_{a}(f)\ainF{\omega}^{a}_{}=(\ainF{\Delta}f)\ainF{l}+(\ainF{D}f)\ainF{n}
-(\bar{\ainF{\delta}}f)\ainF{m}-(\ainF{\delta}f)\bar{\ainF{m}}
\eeq

The necessary equations are the \emph{Cartan's structure equations} (NP-Cartan EDS):
\begin{enumerate}
\item The \emph{Newman-Penrose} (NP) equations
      \beq\label{NP_equations}
      d\ainF{\omega}^{a}+\ainF{\omega}^{a}_{\phantom{1}b}\wedge\ainF{\omega}^{b}=0
      \eeq
      with $\ainF{\omega}^{a}_{\phantom{1}b}$ defining the (linear) matrix valued \emph{connexion} 1-form through the 12 complex 
      \emph{Ricci rotation coefficients} according to
      \beq
      \ainF{\omega}_{ab}=
      \begin{pmatrix}
       0                                          &  \ainF{\omega}_{1}+\bar{\ainF{\omega}}_{1} & -\ainF{\omega}_{2}                         &  -\bar{\ainF{\omega}}_{2} \\
       -\ainF{\omega}_{1}-\bar{\ainF{\omega}}_{1} &  0                                         & \bar{\ainF{\omega}}_{0}                    &  \ainF{\omega}_{0} \\
       \ainF{\omega}_{2}                         &  -\bar{\ainF{\omega}}_{0}                  & 0                                          & -\ainF{\omega}_{1}+\bar{\ainF{\omega}}_{1}\\
      \bar{\ainF{\omega}}_{2}                     & -\ainF{\omega}_{0}                         & \ainF{\omega}_{1}-\bar{\ainF{\omega}}_{1} & 0            
      \end{pmatrix}
      \eeq
      with the allocations
      \bse
      \begin{align}
      &\ainF{\omega}_{0}=\tau\ainF{l}+\kappa\ainF{n}-\rho\ainF{m}-\sigma\bar{\ainF{m}}\\
      &\ainF{\omega}_{1}=\gamma\ainF{l}+\epsilon\ainF{n}-\alpha\ainF{m}-\beta\bar{\ainF{m}}\\
      &\ainF{\omega}_{2}=\nu\ainF{l}+\pi\ainF{n}-\lambda\ainF{m}-\mu\bar{\ainF{m}}
      \end{align}
      \ese 
\item The \emph{Ricci} equations
      \beq\label{Ricci_equations}
      d\ainF{\omega}^{a}_{\phantom{1}b}+\ainF{\omega}^{a}_{\phantom{1}m}\wedge\ainF{\omega}^{m}_{\phantom{1}b}
      =\frac{1}{2}R^{a}_{\phantom{1}bmn}\ainF{\omega}^{m}\wedge\ainF{\omega}^{n}\equiv \ainF{\Omega}^{a}_{\phantom{1}b}
      \eeq
      with $\ainF{\Omega}^{a}_{\phantom{1}b}$ defining the (linear) matrix valued \emph{curvature} 2-form through the 10 complex 
      \emph{curvature scalars} according to
      \beq
      \ainF{\Omega}_{ab}=
      \begin{pmatrix}
       0                                          &  \ainF{\Omega}_{1}+\bar{\ainF{\Omega}}_{1} & -\ainF{\Omega}_{2}                         &  -\bar{\ainF{\Omega}}_{2} \\
       -\ainF{\Omega}_{1}-\bar{\ainF{\Omega}}_{1} &  0                                         & \bar{\ainF{\Omega}}_{0}                    &  \ainF{\Omega}_{0} \\
       \ainF{\Omega}_{2}                         &  -\bar{\ainF{\Omega}}_{0}                  & 0                                          & -\ainF{\Omega}_{1}+\bar{\ainF{\Omega}}_{1}\\
      \bar{\ainF{\Omega}}_{2}                     & -\ainF{\Omega}_{0}                         & \ainF{\Omega}_{1}-\bar{\ainF{\Omega}}_{1} & 0            
      \end{pmatrix}
      \eeq
      with the allocations 
      \bse
      \begin{align}
      \ainF{\Omega}_{0}&=-\Psi_{0}\ainF{n}\wedge\bar{\ainF{m}}-\Psi_{1}(\ainF{l}\wedge\ainF{n}-\ainF{m}\wedge\bar{\ainF{m}})
                        -(\Psi_{2}+2\Lambda)\ainF{m}\wedge\ainF{l}\notag\\
                        &-\Phi_{00}\ainF{n}\wedge\ainF{m}-\Phi_{01}(\ainF{l}\wedge\ainF{n}+\ainF{m}\wedge\bar{\ainF{m}})
                        -\Phi_{02}\bar{\ainF{m}}\wedge\ainF{l}\\
      \ainF{\Omega}_{1}&=-\Psi_{1}\ainF{n}\wedge\bar{\ainF{m}}-(\Psi_{2}-\Lambda)(\ainF{l}\wedge\ainF{n}-\ainF{m}\wedge\bar{\ainF{m}})
                        -\Psi_{3}\ainF{m}\wedge\ainF{l}\notag\\
                        &-\Phi_{10}\ainF{n}\wedge\ainF{m}-\Phi_{11}(\ainF{l}\wedge\ainF{n}+\ainF{m}\wedge\bar{\ainF{m}})
                        -\Phi_{12}\bar{\ainF{m}}\wedge\ainF{l}\\
      \ainF{\Omega}_{2}&=-(\Psi_{2}+2\Lambda)\ainF{n}\wedge\bar{\ainF{m}}-\Psi_{3}(\ainF{l}\wedge\ainF{n}-\ainF{m}\wedge\bar{\ainF{m}})
                        -\Psi_{4}\ainF{m}\wedge\ainF{l}\notag\\
                        &-\Phi_{20}\ainF{n}\wedge\ainF{m}-\Phi_{21}(\ainF{l}\wedge\ainF{n}+\ainF{m}\wedge\bar{\ainF{m}})
                        -\Phi_{22}\bar{\ainF{m}}\wedge\ainF{l}
      \end{align}
      \ese
      where the $\Phi$s correspond to the Ricci components ($\bar{\Phi}_{ij}=\Phi_{ji}$), $\Lambda$ to the Ricci scalar and the $\Psi$s to the Weyl components.
\item The \emph{Eliminant}\cite{Edgar_References} equations (i.e., the $d$ acting on the NP equations)
      \beq
      (d\ainF{\omega}^{a}_{\phantom{1}b}+\ainF{\omega}^{a}_{\phantom{1}m}\wedge\ainF{\omega}^{m}_{\phantom{1}b})\wedge\ainF{\omega}^{b}=0
      \eeq
\item The \emph{Bianchi} equations (i.e., the $d$ acting on the Ricci equations)
       \beq\label{Bianchi_equations}
      d\ainF{\Omega}^{a}_{\phantom{1}b}+\ainF{\omega}^{a}_{\phantom{1}m}\wedge\ainF{\Omega}^{m}_{\phantom{1}b}
      -\ainF{\Omega}^{a}_{\phantom{1}m}\wedge\ainF{\omega}^{m}_{\phantom{1}b}=0
       \eeq
\end{enumerate}
and finally, the six commutator pairs, apparently, correspond to the statement $d(dF)=0$ for any $p$-form $F$.
The section concludes with the following:
\begin{defn}
A $p$-form ($p<4$) in the semi-Riemannian space, is said to constitute an integral of an EDS (here, the NP-Cartan 
system) when its exterior derivative vanishes (i.e., it is closed) \emph{locally} by virtue of the EDS itself.
\end{defn}
The locality is a crucial topological topic to be commented later on.

\section{Finding integrals and identities in the Newman Penrose formalism}
In the present section the focus will be on Kinnersley's initial assumptions, i.e., all the Petrov type D, Ricci 
flat, solutions to the EFEs.

The starting hypothesis implies\cite{Penrose_Rindler} that there exists a family of frames such that $\Lambda=0$, 
$\Phi_{ij}=0$, and $\Psi_{0}=\Psi_{1}=\Psi_{3}=\Psi_{4}=0$ where the last equalities are obtained via the 
implementation of null rotations about first $\ainF{l}$ and second $\ainF{n}$. Thus four of the six real degrees 
of freedom corresponding to the use of the Lorentz group have been used up. Still, there is a residual freedom; 
the system is invariant under spin-boost transformations as per
\beq
(\ainF{l},\ainF{n},\ainF{m},\bar{\ainF{m}}) \rightarrow 
(\zeta\bar{\zeta}\ainF{l},\frac{1}{\zeta\bar{\zeta}}\ainF{n},\frac{\zeta}{\bar{\zeta}}\ainF{m},\frac{\bar{\zeta}}{\zeta}\bar{\ainF{m}})
\eeq
This freedom is the reason for the existence of a whole family of frames.
Based on the transformation laws for the NP variables\cite{Carmeli_Kaye}, it is deduced that the following 
quantities:
\beq
\{\mu\rho, \mu\bar{\rho}, \rho\bar{\mu}, \tau\pi, \tau\bar{\tau}, \pi\bar{\pi}, \Psi_{2}\}
\eeq
are left invariant under the action of the residual freedom. According to the 
\emph{Cartan-Kalrhede}\cite{Cartan_Karlhede} (CK) algorithm all these quantities belong to (but without exhausting) the set of 
the Cartan invariants.

The Bianchi equations \eqref{Bianchi_equations}, along with the NP equations \eqref{NP_equations} imply:
\bse
\begin{align}
&\kappa=\lambda=\nu=\sigma=0\\
& d\Psi_{2}=-3\mu\Psi_{2}\ainF{l}+3\rho\Psi_{2}\ainF{n}+3\pi\Psi_{2}\ainF{m}-3\tau\Psi_{2}\bar{\ainF{m}}\label{first_covariant}
\end{align}
\ese

Now using
\begin{itemize}
\item the NP equations \eqref{NP_equations}
\item the Ricci \underline{component equations} $[1,4]$ and $[2,3]$ of \eqref{Ricci_equations}
\item the fact that $d^{2}\Psi_{2}=0$ (i.e., post Bianchi equation\cite{Edgar_el_al.})
\end{itemize}
four, equations giving $d\mu,d\pi,d\rho,d\tau$ in terms of the spin connexion coefficients emerge: 
\bse
\begin{align}
d\mu&=-(\gamma\mu+\mu^{2}+\mu\bar{\gamma})\ainF{l}+A\ainF{n}+(\alpha\mu+\mu\pi+\mu\bar{\beta}-\pi\bar{\mu})\ainF{m}\notag\\
&+(B+\beta\mu-\gamma\tau+\mu\bar{\alpha}+\tau\bar{\gamma})\bar{\ainF{m}}\\
d\pi&=(-\gamma\pi-\mu\pi+\pi\bar{\gamma}-\mu\bar{\tau})\ainF{l}+F\ainF{n}+(\alpha\pi+\pi^{2}-\pi\bar{\beta})\ainF{m}\notag\\
&+(-A-\epsilon\mu+\beta\pi-\pi\bar{\alpha}-\mu\bar{\epsilon}+\mu\bar{\rho}+\pi\bar{\pi}+\Psi_{2})\bar{\ainF{m}}\\
d\rho&=(\epsilon\rho+\rho^{2}+\rho\bar{\epsilon})\ainF{n}+(-\beta\rho-\rho\tau-\rho\bar{\alpha}+\tau\bar{\rho})\bar{\ainF{m}}\notag\\
&+(-A-\epsilon\mu+\gamma\rho+\rho\bar{\gamma}-\mu\bar{\epsilon}-\tau\bar{\tau}+\pi\bar{\pi})\ainF{l}
+(F-\alpha\rho+\epsilon\pi-\rho\bar{\beta}-\pi\bar{\epsilon})\ainF{m}\\
d\tau&=B\ainF{l}+(\epsilon\tau+\rho\tau-\tau\bar{\epsilon}+\rho\bar{\pi})\ainF{n}+
(-\beta\tau-\tau^{2}+\tau\bar{\alpha})\bar{\ainF{m}}\notag\\
&(A+\epsilon\mu-\alpha\tau+\tau\bar{\beta}+\mu\bar{\epsilon}-\rho\bar{\mu}-\pi\bar{\pi}-\Psi_{2})\ainF{m}
\end{align}
\ese
There are three unknown functions ($A$, $B$, $F$) --reflecting the fact that in each Ricci 2-form two 
directional derivatives, out of the four, are missing. The rest of the Ricci equations involve the operation of 
$d$ on $\alpha, \beta,\gamma, \epsilon$ and will not be needed.

Now one can search for $p$-forms which will be integrals of the NP-Cartan EDS. By construction, and by virtue of the
CK algorithm\cite{Cartan_Karlhede} one can see that for a first, purely algebraic approach to the Petrov type D, Ricci flat spaces, the Riemann tensor (i.e., $\Psi_{2}$) 
and its first covariant derivatives (i.e., $\mu,\pi,\rho,\tau$, --by virtue of the \eqref{first_covariant}), both
being invariant under the residual freedom of spin-boosts (therefore the gauge fixing according to 
the CK algorithm can be performed in the second covariant derivative).
For these reasons the $p$-form to be sought must be left invariant under the residual freedom; thus it may not contain
any of the gauge quantities $\alpha, \beta,\gamma, \epsilon$.

The first attempt will concern the Riemann tensor, for which, a natural candidate is the spin-boost invariant component of
the curvature 2-form
\beq
I_{0}\equiv f(\Psi_{2})\ainF{l}\wedge\ainF{n}+h(\Psi_{2})\ainF{m}\wedge\bar{\ainF{m}}
\eeq
and the property of being an integral reads
\beq\label{zeroth_integral}
d I_{0}|_{\text{EDS}}=0
\eeq
A simple calculation involving both the NP equations\eqref{NP_equations} and the Bianchi equations\eqref{first_covariant} reveals two algebraic integrability conditions 
\bse
\begin{align}
(\tau\bar{\tau}-\pi\bar{\pi})(f(\Psi_{2})+h(\Psi_{2}))=0\\
(\mu\bar{\rho}-\rho\bar{\mu})(f(\Psi_{2})+h(\Psi_{2}))=0
\end{align}
\ese
Thus 
\beq
\Big((\tau\bar{\tau}-\pi\bar{\pi})=0,~~~(\mu\bar{\rho}-\rho\bar{\mu})=0\Big) \vee \Big(f(\Psi_{2})+h(\Psi_{2})=0\Big)
\eeq
i.e., either the desired identities will hold OR $h(\Psi_{2})=-f(\Psi_{2})$. For the time being let 
$h(\Psi_{2})=-f(\Psi_{2})$ only. Then, substitution back to the condition \eqref{zeroth_integral} results in a 
simple ODE for the function $f$, 
\beq
2f(\Psi_{2})-3\Psi_{2}f'(\Psi_{2})=0
\eeq
the solution being
\beq
I_{0}=(\Psi_{2})^{\frac{2}{3}}(\ainF{l}\wedge\ainF{n}-\ainF{m}\wedge\bar{\ainF{m}}), ~~~dI_{0}|_\text{EDS}=0
\eeq
(At this point it should be noted that the same ODE emerges when the identities hold.)
This is a well known result --see e.g., ref.\cite{Israel}. A note on topology is pertinent at this point. Indeed, 
applying the Gauss-Stokes theorem to $I_{0}$ for e.g., the Kerr black hole, one can see that the corresponding
integral assumes a null value if the area of integration does not contain the curvature singularity (at $r=0$ and
$\theta=\pi/2$) (since the integrant is left invariant under continious deformations of the integration area into a point) and it is a constant (related to 
the mass $M$) when the singularity is included --cf. Morera-Cauchy theorems in complex analysis.

The second attempt will concern the first covariant derivative of the Riemann tensor and a natural candidate is the coframe
vector
\beq
I_{1}\equiv \mu\ainF{l}-\rho\ainF{n}-\pi\ainF{m}+\tau\bar{\ainF{m}}
\eeq
which is left spin-boost invariant. By virtue of the Bianchi equations \eqref{Bianchi_equations} the integral
condition is already satisfied,
\beq\label{First_integral}
dI_{1}|_\text{EDS}=0
\eeq
so no new information is gained up to this point and the question about the validity of 
$(\tau\bar{\tau}-\pi\bar{\pi})=0, (\mu\bar{\rho}-\rho\bar{\mu})=0$ \underline{\textbf{is still pending}}.
Never the less ---in order to avoid conceptual degeneracy--- the procedure of finding integrals, must have an end; this end has not been reached yet.\\ Indeed, one can compose new integrals based on the old ones, by using
the Hodge duality and the wedge product. Of cource not all the produced integrals will be functionally independent compared to the old ones.
In the present case, one could consider two more functionally indepenent integrals. The first is the 2-form
\beq
I_{2}\equiv I_{1}\wedge\bar{I_{1}}
\eeq 
Obviously it is
\beq
dI_{2}|_\text{EDS}=0
\eeq
since \eqref{First_integral} holds. Thus
\begin{align}
I_{2}&=(\pi\bar{\mu}+\mu\bar{\tau})\ainF{l}\wedge\ainF{m}+(-\tau\bar{\mu}-\mu\bar{\pi})\ainF{l}\wedge\bar{\ainF{m}}
+(\rho\bar{\mu}-\mu\bar{\rho})\ainF{l}\wedge\ainF{n}\notag\\
&
+(-\tau\bar{\tau}+\pi\bar{\pi})\ainF{m}\wedge\bar{\ainF{m}}+(\pi\bar{\rho}+\rho\bar{\tau})\ainF{m}\wedge\ainF{n}
+(-\tau\bar{\rho}-\rho\bar{\pi})\bar{\ainF{m}}\wedge\ainF{n}
\end{align}
\underline{\textbf{This integral is new to the literature}} --at least to the best of the author's knowledge. The reader must 
have observed that the identities to be proven are involved in this integral. But one must also consider the last, 
functionally independent integral, i.e. the 0-form,
\beq
I_{3}\equiv\star{(I_{1}\wedge I_{2}\wedge\bar{I_{2}})}=(\rho\bar{\mu}-\mu\bar{\rho}+\tau\bar{\tau}-\pi\bar{\pi})(\Psi_{2})^{\frac{2}{3}}
\eeq
where the $\star$ denotes the Hodge dual. On one hand this is an integral for it is constructed by other integrals.
On the other hand its exterior derivative contains the coefficients $\alpha, \beta,\gamma, \epsilon$ --which do not 
behave under the residual freedom. The only logical resolution to this puzzle is given by the constraint
\beq
I_{3}=0
\eeq
which holds only if both its real and imaginaty parts vanish, i.e., if and only if
\beq
\rho\bar{\mu}-\mu\bar{\rho}=0, ~~~\tau\bar{\tau}-\pi\bar{\pi}=0.
\eeq

The logical circle now closes. Three integrals for the NP-Cartan EDS system have been found and two identities 
emerged naturally as integrability conditions. To recapitulate, the three integrals are:
\bse
\begin{align}
I_{0}&=\Psi^{\frac{2}{3}}(\ainF{l}\wedge\ainF{n}-\ainF{m}\wedge\bar{\ainF{m}}), ~~~dI_{0}|_\text{EDS}=0\\
I_{1}&=\mu\ainF{l}-\rho\ainF{n}-\pi\ainF{m}+\tau\bar{\ainF{m}}, ~~~dI_{1}|_\text{EDS}=0\\
I_{2}&=(\pi\bar{\mu}+\mu\bar{\tau})\ainF{l}\wedge\ainF{m}+(-\tau\bar{\mu}-\mu\bar{\pi})\ainF{l}\wedge\bar{\ainF{m}}\notag\\
&+(\pi\bar{\rho}+\rho\bar{\tau})\ainF{m}\wedge\ainF{n}
+(-\tau\bar{\rho}-\rho\bar{\pi})\bar{\ainF{m}}\wedge\ainF{n}, ~~~dI_{2}|_\text{EDS}=0
\end{align}
\ese
and as a concequence of the termination for the procedure of generating all the functionally independent integrals:
\beq
\rho\bar{\mu}-\mu\bar{\rho}=0, ~~~\tau\bar{\tau}-\pi\bar{\pi}=0.
\eeq

\end{document}